# Bottom-up Growth of Graphene Nanospears and Nanoribbons

*Haibin Sun[1,2,+], Fengning Liu[1,3,+], Leining Zhang[1+], Ben McLean,[1] Hao An,[1,3] Ming Huang,[1] Marc-Georg Willinger,[4] Rodney Ruoff,[1,3,5,6] Zhujun Wang,[4,7,*] and Feng Ding[1,3,*]*

[1]Center for Multidimensional Carbon Materials, Institute for Basic Science, Ulsan 44919, Republic of Korea
[2]Key Laboratory of Microelectronics and Energy of Henan Province, College of Physics and Electronic Engineering, Xinyang Normal University, Xinyang 464000, China
[3]Department of Materials Science and Engineering, Ulsan National Institute of Science and Technology, Ulsan 44919, Republic of Korea
[4]Scientific Center for Optical and Electron Microscopy, ETH Zurich, Otto-Stern-Weg 3, 8093 Zurich, Switzerland
[5]Department of Chemistry, Ulsan National Institute of Science and Technology (UNIST), Ulsan, 44919, Republic of Korea
[6]School of Energy and Chemical Engineering, Ulsan National Institute of Science and Technology (UNIST), Ulsan, 44919, Republic of Korea
[7]School of Physical Science and Technology, Shanghai Tech University, Shanghai 200031, China
[+] These authors contributed equally to this work.

[*]*email:* wangzhj3@shanghaitech.edu.cn, f.ding@unist.ac.kr

ABSTRACT

Graphene nanoribbons (GNRs) are considered one of the most promising materials for next generation electronics, however a reliable and controllable synthesis method is still lacking. Here, we report the CVD growth of GNRs on a copper surface and the corresponding mechanisms of growth. One-dimensional GNR growth is enabled by a vapor-liquid-solid (VLS) graphene growth guided by on-surface propagation of a liquid catalyst particle. Controlling the suppression of vapor-solid-solid (VSS) graphene growth along the width direction of the GNR by tuning the flow of H2 during growth gives rise to a spear head-shaped graphene that we term graphene nanospears (GNSs). The real-time visual and spatially resolved observations confirm the VSS growth of graphene can be fully suppressed and lead to GNR formation on Cu surface. These findings reveal key insight into the growth mechanism of graphene and open a door for achieving a facile and scalable method of synthesizing free standing GNRs.

1. Introduction

Graphene nanoribbons (GNRs), inherit properties from large two-dimensional (2D) graphene such as high mobility,[1, 2] high mechanical,[3, 4] and thermal stability.[5, 6] Notably, GNRs exhibit a band gap which can be tuned depending on the GNR width and its crystallographic orientation,[7-13] and therefore offer exciting opportunities for application in electronics.[14-17] However, applications of GNRs is greatly hindered by the challenge of their difficult synthesis.

Top-down methods of GNRs synthesis, such as unzipping carbon nanotubes [18-21] and etching graphene,[22-25] have proven challenging, as they commonly introduce defects and irregular edge structures at low yields with difficulty in controlling GNR width. For the last decade,[26] bottom-up synthesis of GNRs from aromatic molecular frameworks has seen more success, despite often requiring costly ultra-high vacuum (UHV) conditions,[27] and those based in solution are difficult to transfer to insulating substrates and edge functionalization can diminish favorable optoelectronic properties.[28, 29] Chemical vapor deposition (CVD) presents a facile alternative to bottom-up GNR synthesis and has been utilized to grow GNRs on Au(111) surfaces[30-32] though often have imperfect edges and short lengths. Cu(111) provides a more accessible substrate for GNR synthesis though has only been successful under UHV conditions.[33-37] Recently, controlled growth of oriented GNRs *via* CVD was achieved in etched hBN channels on a quartz substrate,[38, 39] and suspended GNR arrays were synthesized in high yield on Ni following plasma CVD and cooling,[40] however the mechanisms of growth for each case lack understanding.

Here a unique approach to CVD growth of GNRs on a Cu surface is reported. Atmospheric pressure CVD graphene growth yields graphene islands exhibiting a variety of shapes including the formation of what we term graphene nanospears (GNSs). The mechanism of GNSs growth is different from CVD graphene [41-43] that is typically grown *via* a vapor-solid-solid (VSS) mechanism, where individual graphene islands nucleate and grow from precursor decomposition and addition to the graphene edge on the Cu surface.[44-46] The observed sharp GNS structures form as a result of a suppression of the VSS growth of graphene in favor of vapor-liquid-solid (VLS) [47] growth of a central GNR structure. By controlling $H_2$ flow in a flexible *in-situ* SEM system, the complete suppression of VSS growth leads to the VLS-grown GNRs. In other words, two-dimensional VSS graphene growth is degenerated in favor of one-dimensional VLS GNR growth. This study reveals that VLS growth of GNRs is possible, paving a way for large-scale synthesis of free standing GNRs.

## 2. Results and Discussion

Performing atmospheric pressure CVD at 1050 °C, with 150 sccm of diluted $CH_4$ and 200 sccm of $H_2$ (**Figure. S1a**) we observe graphene islands presented in **Figure. 1(a-h)** and **Figure. S2-3**. Consistent with previous graphene growth on Cu foils, a large number of graphene islands are hexagonal shaped (Figure. 1b). Interestingly, graphene islands exhibiting a range of other shapes are also observed. Figure. 1c-h shows scanning electron microscopy (SEM) images of graphene islands with elongated, sharp tips that terminate in a particle. Graphene islands with one or more particles attached to the tips take the shape of various irregular polygons. As shown in Figure. 1e, a parallelogram-shaped island is present when there are two particles attached, with the two sharp tips having tip angles of ~ 60°. Tip angles ranging from 13-60° are observed. Sharp tip angles are only observed in the presence of a particle at the tip, making it clear that these particles play a crucial role during the growth process of these graphene islands with sharp tips. Figure. 1i-k presents the atomic force microscopy (AFM) image of a parallelogram-shaped graphene island and the magnified images of its two sharp tips, with particles clearly attached. Figure. 1l is the Energy-dispersive X-ray spectroscopy (EDS) analysis of the particle, implying significant Si and O contents in the particle (X-ray photoelectron spectroscopy (XPS) spectra displaying a strong Si-2p signal in **Figure. S4**). Based on the observed coalescence behavior of these particles (will be presented in **Figure. 4**), we can deduce that these particles are in a liquid state with a melting point lower than the experimental temperature (1050 °C). Therefore, these particles are notably not SiO$_2$ particles (melting point 1710 °C) as observed in most CVD experiments for graphene growth. In addition, from the phase diagram of Cu-Si,[48, 49] we find that a Cu$_{(1-x)}$Si$_x$ alloy with x from 0.15 to 0.3 has the melting point of ~ 850 °C. It is reasonable to guess that the particle is then a Cu-rich liquid Cu-Si alloy although the exact content cannot be determined in this system.

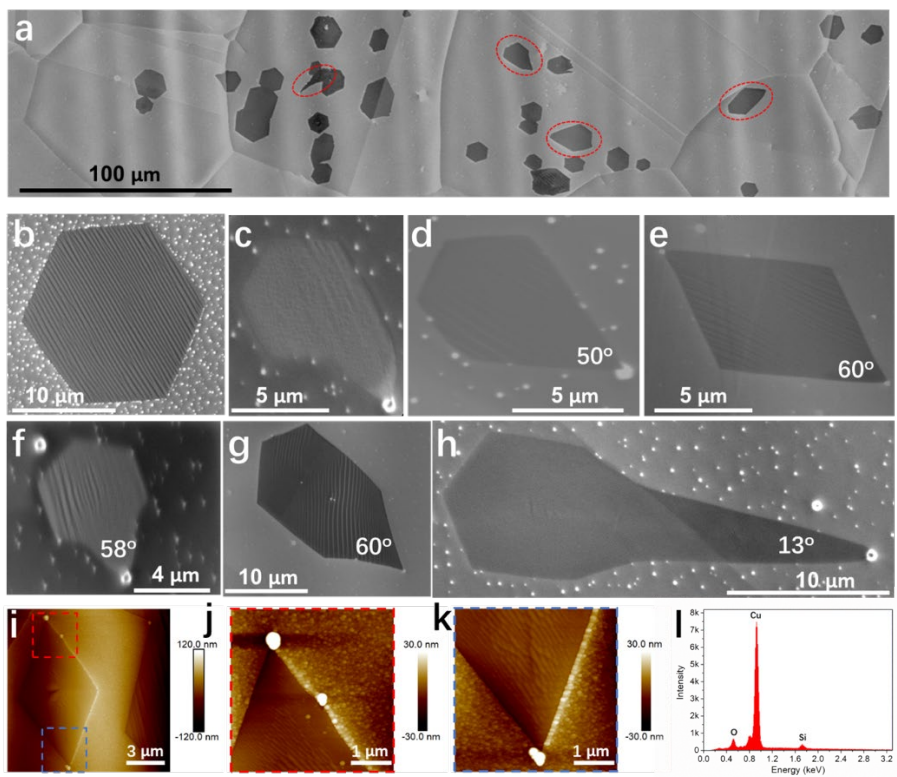

**Figure 1. Catalyst particle-mediated graphene islands grown on Cu foil.** (a) SEM image of graphene islands with different shapes; (b) hexagonal graphene island typically observed following CVD growth; (c-h) catalyst-mediated graphene growth resulting in various graphene island shapes with particles (white) attached to the tips and many particles present on Cu foil; (i-k) AFM images of a parallelogram-shaped graphene island grown on Cu and the magnified zones with catalyst particles; (l) EDS spectrum of the catalyst particle, showing strong Si and O signals.

On a Cu surface, the propagation of graphene edge is the consequence of adding carbon atoms from vapor phase, hence the VSS growth mechanism. The typical hexagonal shape of graphene islands is due to anisotropic growth of the graphene lattice with a six-fold symmetry. As a result of the slow VSS propagation of the zigzag (ZZ) edge, each side of the hexagonal island comprises of ZZ edges.[50-53] According to the crystal growth kinetics, the formation of a sharp-tipped graphene island requires a faster growth rate along the direction of the bisect angle of the tip.[45, 53, 54] Here each sharp tip corresponds to the presence of a particle hence the particles are responsible for the faster growth rate and the symmetry breaking of graphene growth.

To understand how a particle facilities the graphene growth, we recall the VLS growth mechanism of nanowires and carbon nanotubes.[55-59] In this model, feedstock from the vapor phase adsorbs to the liquid phase catalyst particle which facilitates decomposition of the precursors and subsequent attachment to the solid nanowire or nanotube. Recently, the VLS growth of 2D materials has been observed in the growth of ultralong $MoS_2$ nanoribbons.[60, 61] Here we use the combination of the liquid catalyst-guided VLS GNR growth and the general VSS graphene growth to explain the formation of the various shapes of graphene islands observed in Figure. 1. As shown in **Figure. 2a**, the propagation of a liquid particle on Cu surface leads to the formation of a GNR behind it (the black ribbon) while simultaneously, slower VSS growth of the GNR edge leads to a surrounding graphene layer (the gray area). The angle, $\beta$ of the spear-like tip is

$$\beta = 2\sin^{-1}\left(\frac{R_{VSS}}{R_{VLS}}\right) \tag{1}$$

where $R_{VSS}$ and $R_{VLS}$ are the VSS and VLS growth rates of graphene, respectively. From Equation (1), we can clearly see that the shape of the graphene nanospear can be tuned by varying $R_{VSS}$, $R_{VLS}$ or both. As shown in Figure. 2b, a sharp graphene nanospear can be formed if $R_{VLS} \gg R_{VSS}$. Critically, if the VSS graphene growth is completely suppressed, a GNR will be formed by catalyst guided VLS growth (Figure. 2c).

Following the proposed mechanism, we illustrate the growth process of a few graphene islands guided by the propagation of the catalyst particles. Figure. 2d1-d4 corresponds to the growth mechanism of the graphene island in Figure. 1f-h; Figure. 2e1-e4 present the growth process of the graphene island in Figure. 1c-d; Figure. 2f1-f4 predicts how the parallelogram in Figure. 1e may form. The observed graphene islands have all formed as a result of the proposed VSS + VLS mechanism, with the liquid catalyst particle propagating across the Cu surface.

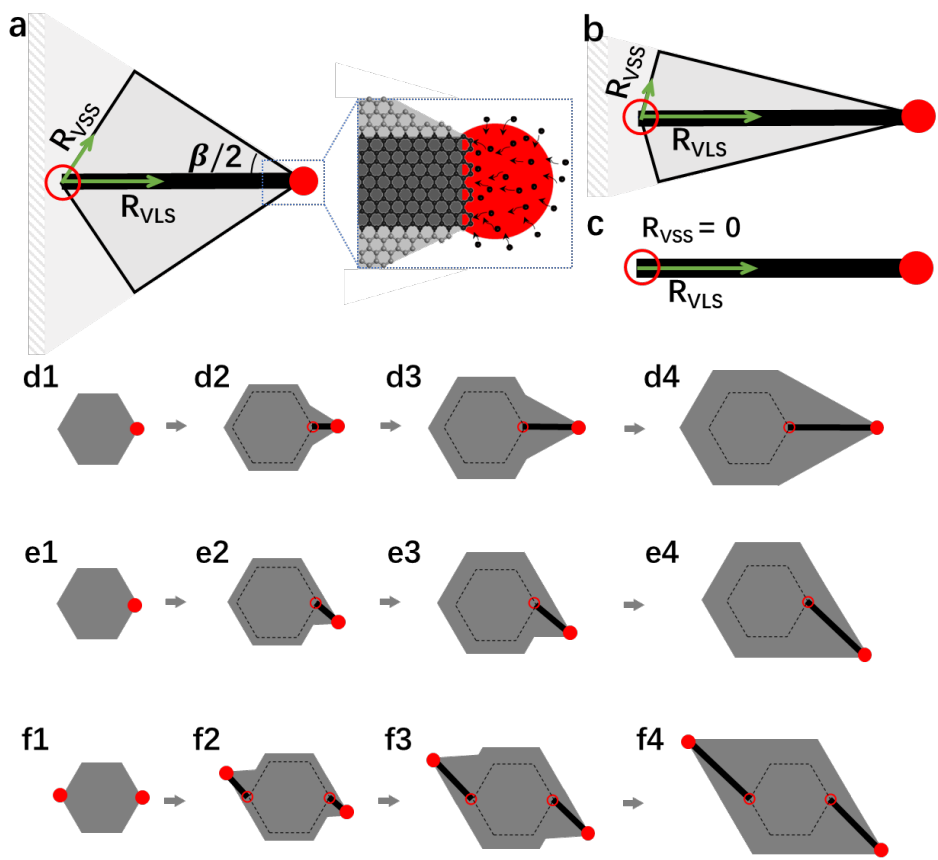

**Figure 2. The mechanism of catalyst particle mediated graphene growth.** (a) A liquid catalyst particle (red circle) moves forward at the rate of $R_{VLS}$ initiating continuous vapor-liquid-solid (VLS) growth of GNR (black line) behind it. The normal VSS growth of graphene (gray area) on Cu surface leads to a GNS with tip angle of β, where $R_{VSS}$ is the graphene growth rate on Cu foil ($R_{VLS} > R_{VSS}$). (b) Slower VSS graphene growth compared to the VLS GNR growth leads to a sharper GNS ($R_{VLS} \gg R_{VSS}$). (c) Suppressing the VSS graphene growth completely ($R_{VSS} = 0$) leads to only GNR growth. (d1 → d2 → d3 → d4) A liquid catalyst particle attached to a hexagonal graphene island initiates VLS GNR growth towards the right and leads to an elongated hexagon with a sharp tip. (e1 → e2 → e3 → e4) A liquid catalyst particle attached to a hexagonal graphene island initiates diagonal GNR growth and leads to an elongated pentagon with a sharp tip. (f1 → f2 → f3 → f4) Two liquid catalysts particles attached to a hexagonal graphene island initiate the growth of two GNRs diagonally leads to a parallelogram-shaped graphene island with two sharp tips.

We demonstrate that the shape of the graphene nanospear can be tuned by varying the experimental condition. By increasing the $H_2/CH_4$ ratio, which has previously been shown to facilitate the etching of graphene and slow down the VSS growth rate,[25, 62] we have successfully synthesized very sharp GNSs, comprising of a central GNR (**Figure. 3a-b** and **Figure. S5**). The SEM images of a typical GNS with a tip angle of 10° is shown in Figure. 3c-d. This tip angle implies that the VLS growth rate is ~ 10 times that of the VSS growth rate. In the central part of Figure. 3d, a narrow ribbon-like track can be clearly seen. After it was transferred to a $SiO_2$ surface, a dark line of ~ 200 nm wide is notable (Figure. 3e), indicative of a GNR, supporting the proposed two-step mechanism shown in Figure. 2. The central GNR is grown *via* the VLS mode guided by the liquid catalyst particle while the surrounding graphene is grown *via* the VSS mode. The AFM image (Figure. 3f) confirms that the central GNR is different from the surrounding graphene. Figure. 3g measures the size of the catalyst particle to be 100 nm in height and the GNR to be 50 nm in width. Furthermore, statistical result shown in Figure 3(h) presents that the tip angles of these GNSs are distributed in a small range between 1 to 12º, implying that increase $H_2/CH_4$ ratio can greatly increase the ratio of $R_{VLS}$ to $R_{VSS}$ by suppressing the VSS graphene growth. The growth mechanism of the central VLS GNR is illustrated in Figure 3(i), where the precursor molecules ($CH_4$) decompose on the Cu substrate and the released C atoms diffuse to the liquid catalyst particle. The liquid catalyst particle facilitates the attachment of C atoms to an end of central GNR by overcoming a lower reaction barrier than that of VSS graphene growth.

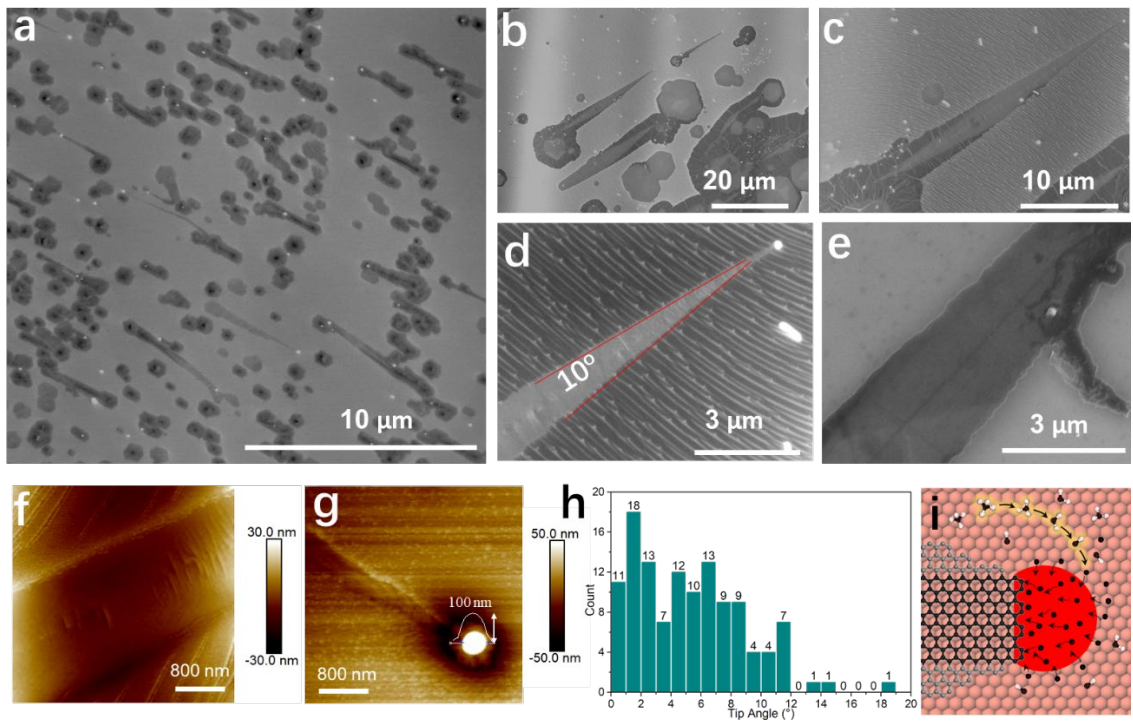

**Figure 3. The growth of sharp GNSs comprising GNRs on Cu foil.** (a) SEM image showing approximately 20% of graphene islands grow as GNSs after increasing the $H_2$ flow rate from 200 sccm to 300 sccm. (b-d) GNSs with different magnifications by SEM. (d) Magnified SEM image showing the tip angle of this GNS is only 10°. (e) Magnified SEM image of a GNS transferred to $SiO_2$ surface clearly showed a central GNR within the GNS. (f-g) AFM images of the GNS wall and tip, within which the central GNR can be seen. Particle measured to be ~ 100 nm in diameter. (h) Statistical results showing the distributions of tip angles. (i) A schematic model showing the growth mechanism of the central VLS GNR and GNS.

The observations in Figure. 3 support our hypothesis that the growth of a GNS contains a VLS-grown GNR central to the VSS-grown graphene edges. However, our thermal CVD system is incapable of reliably tuning the experimental condition to fully suppress VSS growth of graphene surrounding the GNR. To realize the VLS growth of the GNR while suppressing the VSS growth, we employ a more flexible *in-situ* SEM system. **Figure. 4** (a1→ a2 → a3 → a4) demonstrates the real-time and real-space growth of a GNR *via in-situ* SEM. The GNR grows in length by approximately 120 nm/s with no significant change in width. During the whole growth process, the width of the GNR remains ~ 500 nm, implying that in-plane VSS growth is completely suppressed. In turn, this suggests that the activation energy of graphene VSS growth is much higher than that of GNR VLS growth facilitated by the catalyst particle.

Figure. 4 (b1→ b2→ b3→ b4→b5) presents another example of GNR growth by *in-situ* SEM, during which the catalyst particle becomes smaller and smaller and finally disappears from view. As the catalyst particle appears to shrink, the growth of the GNR slows and the width gradually decreases until growth ceases at 104.5 s after the catalyst disappears. This observation highlights the critical role of the catalyst particle in growth, in support of the VLS growth mechanism.

Figure. 4 (c1→ c2→ c3→ c4) demonstrates the liquid nature of the catalyst particles on the Cu surface under the growth condition. During this evolution, it can be clearly seen that particles 1, 2, and 3 merge into a large particle after 72 s; while particle 4, increasing gradually in size, diffuses towards the newly formed particle over 108 s. The quick coalescence and diffusion of ~100 nm diameter catalyst particles strongly support the liquid state of these particles, as discussed above.

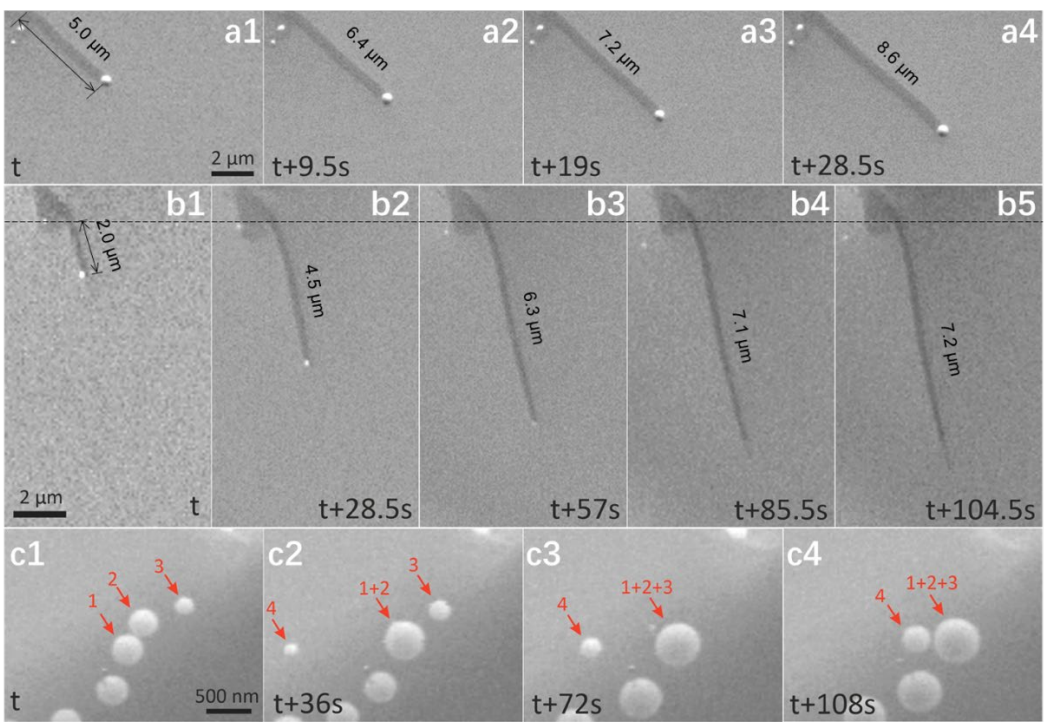

**Figure 4. *In situ* observation of VLS GNR growth on Cu foil.** (a1→ a2→ a3→a4), (b1→b2→b3) The linear growth of two GNRs. (b4→b5) GNR growth ceasing after catalyst dissolution into the Cu foil. (c1→c2→c3→c4) Coalescence of liquid catalyst particles.

## 3. Conclusion

The growth of GNSs that comprise a GNR by atmospheric pressure CVD has been reported, utilizing the propagation of liquid catalyst particles on a Cu surface. Controlling the $H_2$ flow during growth allows for the rates of VSS graphene growth and VLS GNR growth to be tuned. In turn, this controls the tip angle hence length and width of the GNS and potentially of the GNR. Controlling the dimensions along the GNS tip allows for selectivity regarding electronic properties, as the band gap of the material will depend on the width. Complete suppression of VSS graphene growth was demonstrated by *in situ* SEM to preferentially grow a free standing GNR on Cu foil, previously only achievable under UHV conditions. Achieving this level of control in CVD creates the first facile method under atmospheric conditions of synthesizing free standing GNRs.

## 4. Methods

*Atmospheric pressure CVD growth of graphene nanospears on Cu foils.* Under atmospheric pressure (760 Torr), a polycrystalline Cu foil (Alpha Aesar#46365, 25 μm, 99.8 wt.%) was heated to 1050 °C in purified Ar gas (500 standard cubic centimeters per minute (sccm)) and then annealed for 60 min at this temperature in a mixed gas (500 sccm Ar, 200 sccm $H_2$). Next, diluted (1%) in Ar $CH_4$ (150 sccm) was added into the constant mixed gas for 60 min. Finally, the sample was rapidly cooled to room temperature. The temperature-time profile is shown in Figure. S1a.

By tuning the experimental condition, GNSs with very sharp tips were achieved. As shown in Figure. S1b, the Cu foil was heated to 1000 °C in mixed gas (500sccm Ar, 100 sccm $H_2$) and then annealed for 60 min. After that, purified $CH_4$ (5 sccm) and $H_2$ (300 sccm) was introduced the CVD chamber for 10 min. Finally, the sample was quickly cooled to room temperature.

*In-situ observations by environmental SEM.* *In-situ* growth of GNR was performed inside the chamber of a modified environmental SEM system (Thermo Fisher Quattro S) with a custom infrared laser heating stage and with gas supplied station. The growth temperature arranged from 900 °C to 1000 °C under 40 Pa. Images were recorded by a large-field detector. During the experiments, the microscope was operated at an acceleration voltage of 3 kV. No influence of the electron beam on the growth and etching process could be observed. The imaged regions and their respective surroundings showed similar behavior, as evidenced by changing the magnification or by moving the sample under the beam. Furthermore, no electron beam induced contamination was observed at elevated temperatures.

*Characterization.* Optical images were obtained using an Olympus BX51optical microscope. SEM images were conducted with an FEI Verios 460 ESEM field emission gun at 20 keV. AFM data was performed with a Bruker Dimension Icon system. Raman spectra and maps were performed with a WITec scope with a 532 nm wavelength excitation Ar laser.

**Supporting Information**

Supporting Information is available free of charge.

ASSOCIATED CONTENT

AUTHOR INFORMATION

**Corresponding Author**

*Feng Ding* − Center for Multidimensional Carbon Materials, Institute for Basic Science (IBS), Ulsan 44919, Republic of Korea; Department of Materials Science and Engineering, Ulsan National Institute of Science and Technology (UNIST), Ulsan 44919, Republic of Korea.

*Zhujun Wang* − School of Physical Science and Technology, Shanghai Tech University, Shanghai 200031, China**Author Contributions**

F. Ding conceived and designed the experiments and calculations. H. Sun and F. Liu performed the synthesis and structural characterization. Z. Wang and M. Willinger performed *in situ* SEM experiments. L. Zhang carried out the calculations and data analysis. B. McLean, H. An, M. Huang, R. Ruoff took part in the experimental analysis. F. Ding, H. Sun, B. McLean, and L. Zhang wrote the manuscript. All authors reviewed the manuscript and approved the submission.

**Notes**

The authors declare no competing financial interest.

ACKNOWLEDGMENT

The authors acknowledge support from the Institute for Basic Science (IBS-R019-D1) of Republic of Korea.

TOC

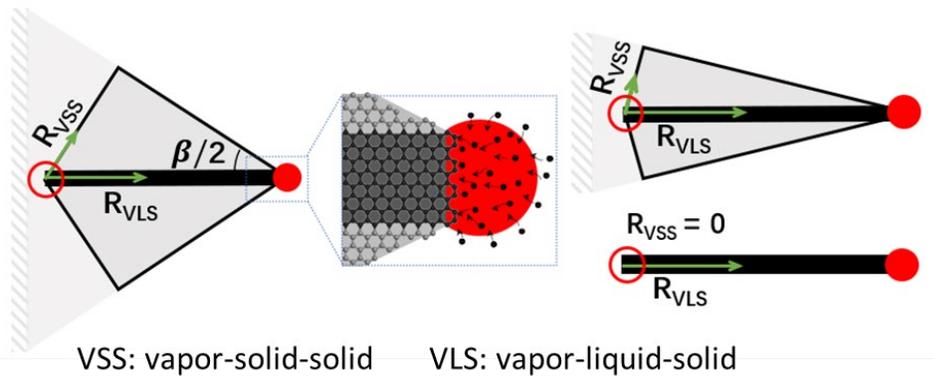

VSS: vapor-solid-solid    VLS: vapor-liquid-solid

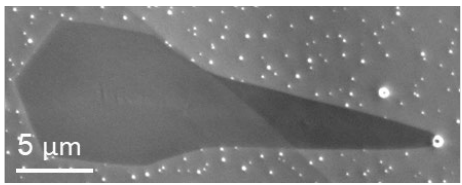
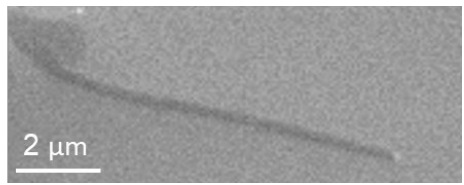